\documentclass[journal]{IEEEtran}
\usepackage{cite}
\usepackage{amsmath,amssymb,amsfonts}
\usepackage{algorithm,algorithmic}
\usepackage{graphicx}
\usepackage{textcomp}
\usepackage{xcolor}
\usepackage[caption=false]{subfig}
\usepackage{pifont}
\usepackage{multirow}
\usepackage{makecell}
\usepackage{bookmark,booktabs}

\ifCLASSINFOpdf
\else
\fi

\begin{document}

\title{HACache: Leveraging Read Performance with Cache in a Heterogeneous Array}


\author{Jialin Liu,  Liang Shi, Dingcui Yu

\thanks{
Jialin Liu, Liang Shi and Dingcui Yu are with the School of Computer Science and Technology, East China Normal University. 
Emails: 52255901007@stu.ecnu.edu.cn, shi.liang.hk@gmail.com, dingcuiy@gmail.com
The corresponding author is Liang Shi (Emails: shi.liang.hk@gmail.com).}
}


\maketitle

\begin{abstract}
In cost-sensitive deployments, RAID arrays may combine SSDs with different performance levels. 
Such heterogeneity arises when aging SSDs degrade yet remain usable, or when failed drives are replaced with new devices of explicitly better performance. 
While this reduces procurement cost, it creates performance challenges:
traditional striping mecahnism distributes requests evenly, but slower SSDs become bottlenecks, leaving faster ones underutilized and limiting overall bandwidth to the slowest drive.

To address this, we propose \textbf{HACache (Heterogeneity Adaptive Cache)} for read-intensive workloads. HACache introduces high-performance SSDs as read caches to rebalance request distribution. 
First, we formalize the request diversion problem and solve it formally.
Second, to support optimal diversion ratios searching at runtime, HACache adopts a two-phase request diversion ratio adjustment mechanism.
Finally, a cache capacity regulation is adopted to adapt quotas for each backend SSD based on hit rates and request diversion needs.
This design maximizes bandwidth utilization. 
Experiments show HACache improves heterogeneous RAID read performance significantly, with bandwidth gains of about 35\% in typical mixed configurations.

\end{abstract}


\IEEEpeerreviewmaketitle

\section{Introduction}
RAID systems are designed to combine multiple SSDs into a single logical unit for aggregated performance and data protection.
Although RAID was originally designed for HDDs, recent advances in SSD technology have spurred growing interest in SSD‑based RAID across both research and commercial products.
Compared with HDD-based RAID, SSD-based RAID systems can deliver much higher IOPS and lower latency, making it suitable for latency‑sensitive and high‑throughput workloads.
Also, with the emerging high-capacity SSDs, SSD-based arrays can provide higher storage density than HDD-based arrays.

In cost-constrained deployment scenarios, storage systems may need to construct RAID arrays by mixing SSDs with different performance specifications.
The causes of such heterogeneous configurations are diverse.
On the one hand, after long-term operation, some large-capacity SSDs may suffer performance degradation due to aging but remain usable;
On the other hand, during array operation, some SSDs may fail and be replaced with newly purchased devices of different specifications.
Although large data centers and cloud providers typically address such issues by replacing drives until homogeneity is restored, such replacements are prohibitively expensive in cost‑constrained scenarios.
However, maintaining such heterogeneous configurations may introduce challenges in utilization since traditional RAID systems are mainly designed for homogeneous configurations.

Under traditional RAID architectures, the striping mechanism distributes requests approximately evenly across SSDs.
But under heterogeneous RAID architectures, processing speeds of SSDs may differ significantly from each other's.
This mismatch between request distribution and processing speed causes faster SSDs to be slowed down by slower ones.
Specifically, when backend SSDs exhibit performance disparities, slower SSDs accumulate large request queues, while faster SSDs often complete tasks early and become idle, leading to insufficient overall bandwidth utilization.
This problem directly results in each SSD in a heterogeneous array being limited to the minimum bandwidth among all SSDs and causes serious bandwidth underutilization.
Previous works on heterogeneous arrays \cite{adaptraid,asymmetricraid,raidx} mainly focus on solving the capacity heterogeneity and hardly solve the above mentioned bandwidth underutilization problem

Focusing on read-intensive application scenarios, this paper proposes HACache (Heterogeneity Adaptive Cache), a cache mechanism that is aware of performance differences.
We introduce high-performance SSDs into the array as read caches to adjust the proportion of requests allocated to each drive.
First, to address the mismatch between request distribution and processing speed, HACache provides an optimization objective for cache diversion ratios to determine how requests are split across SSDs.
Second, to enable dynamic optimization based on this objective, HACache offers a runtime strategy that automatically searches for optimal diversion ratios.
Finally, to fully utilize cache space, HACache provides a cache capacity regulation strategy that dynamically adjusts each SSD’s cache quota based on diversion conditions and cache hit capability.
This design maximizes the utilization of overall array bandwidth.

Experimental results show that the mechanism significantly improves read performance in heterogeneous RAID.
In typical mixed configurations, the system’s aggregate bandwidth improves by about 35\% compared with baseline schemes.
These results validate the effectiveness of our design and demonstrate that, in heterogeneous SSD environments, performance-aware caching strategies can achieve a balance between cost and performance.

\section{Background and Motivation}
\subsection{Heterogeneity in Large-Capacity SSD Arrays}

In the initial deployment stage of storage arrays, systems are typically configured with homogeneous SSDs to ensure balanced global performance.
However, as system runtime increases, individual devices within the array inevitably experience varying degrees of performance degradation or hardware replacement, leading to significant disparities among devices. 
We formally define this evolutionary phenomenon as \textbf{array heterogeneity}.

In practical production environments, the root causes of array heterogeneity are diverse.
This section focuses on two representative scenarios:

The first scenario arises when certain SSDs operate in a degraded mode, where device read/write bandwidth drops significantly but basic data correctness and availability remain intact.
Common causes include flash wear-out and localized hardware anomalies~\cite{failslow,capvar}.
Particularly with the trend toward ultra-large SSDs, existing studies~\cite{reparo} propose improving reliability by limiting the fault domain: when individual flash dies suffer permanent damage, the system masks the failed die at the physical layer and relies on array-level redundancy to rebuild local data.
Under this mechanism, the logical functionality of the damaged SSD is preserved, but the reduction in internal parallelism inevitably leads to degraded read performance.
Given the high cost of large-capacity SSDs, replacing an entire device due to partial degradation incurs prohibitive expense; retaining such drives in service becomes a direct cause of performance heterogeneity within the array.

The second scenario stems from complete hardware failures and generational upgrades.
After years of operation, some SSDs fail completely and must be physically replaced.
Due to rapid semiconductor evolution, earlier models may already be discontinued, forcing administrators to introduce newer-generation SSDs.
Unlike the relatively gradual performance evolution of HDDs, the performance gap between SSD generations is dramatic. For example, SSDs with PCIe 4.0 interfaces achieve peak read bandwidths of about 7 GB/s, while PCIe 5.0 devices reach 13–14 GB/s.
This “mixed old and new” configuration caused by passive hardware upgrades constitutes another typical heterogeneous array scenario.

In theory, heterogeneity in read performance, write performance, and capacity can all affect arrays. This chapter focuses on read-intensive workloads such as dataset loading, content delivery networks, and big data analytics. Accordingly, subsequent architectural design and mechanism discussions will concentrate on the problem of \textbf{read performance heterogeneity}.

\subsection{Bandwidth Underutilization in Heterogeneous Arrays}
\subsubsection{Phenomenon Description}
\label{sec:c4:motiv:phenomenon}

\begin{figure}[htb]
    \centering
    \subfloat[4KB Random Read]{
        \label{fig:c4:motiv:bw1}
        \includegraphics[width=0.45\columnwidth]{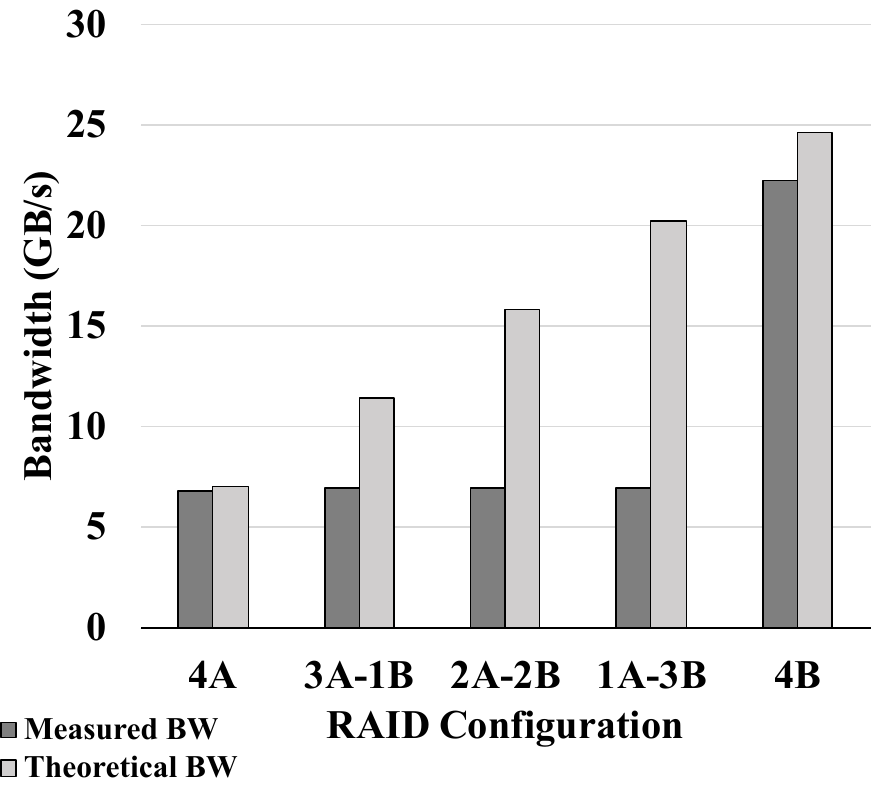} 
    }
    \subfloat[128KB Sequential Read]{
        \label{fig:c4:motiv:bw2}
        \includegraphics[width=0.45\columnwidth]{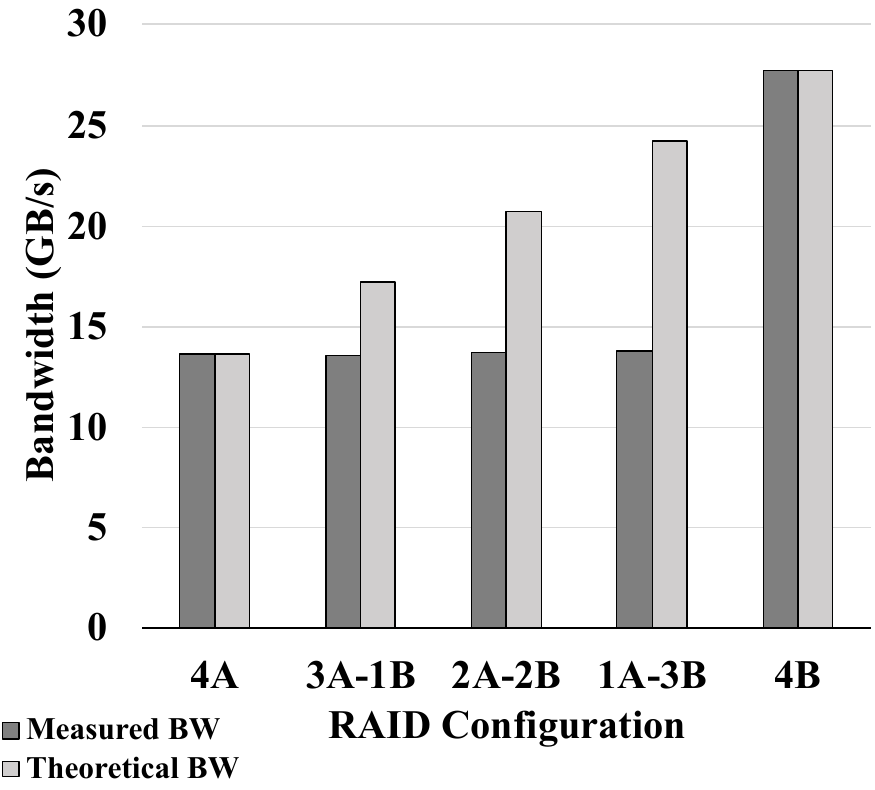}
    }
    \caption{Measured vs. theoretical bandwidth of 4‑drive heterogeneous RAID}
 \label{fig:c4:motiv:bw}
\end{figure}

To quantitatively evaluate the impact of read‑performance heterogeneity on storage arrays, we conducted real‑system benchmarks on a RAID‑5F array~\cite{spdk} composed of two types of SSDs. In the following discussion, devices with higher throughput are referred to as \emph{fast drives}, while those with limited throughput are referred to as \emph{slow drives}.

The performance parameters of the physical devices are listed in Table~\ref{tab:c4:eval:ssds}. In our experiments, type‑A devices serve as slow drives and type‑B devices as fast drives (other platform configurations are detailed in Section~\ref{sec:c4:eval:setup}). We applied random read workloads with 4KB requests and sequential read workloads with 128KB requests, with the stripe size fixed at 128KB. In addition, we calculated the ideal aggregate bandwidth by summing the theoretical bandwidths of all member drives, and plotted both theoretical and measured throughput in Figure~\ref{fig:c4:motiv:bw}. The x‑axis indicates the specific array topology of four SSDs (e.g., “3A‑1B” denotes three type‑A slow drives and one type‑B fast drive).

Based on the benchmark results, we derive the following key observations:

\begin{itemize}
\item \textbf{Ideal scalability of homogeneous arrays:} In purely homogeneous configurations (“4A” or “4B”), the measured bandwidth nearly reaches the theoretical maximum. Except for the “4B” array under 4KB random reads, which is limited by host CPU concurrency and achieves 90.2\% of the theoretical value, all other homogeneous groups achieve at least 97\% of the theoretical bandwidth.
\item \textbf{Bandwidth underutilization in heterogeneous arrays:} Once heterogeneity is introduced, measured throughput drops sharply, far below the theoretical aggregate. This reveals the fundamental problem of \textbf{bandwidth underutilization}. The issue is most severe in the “1A‑3B” configuration under 4KB random reads, where measured bandwidth is only 34\% of the theoretical value; other heterogeneous configurations also fail to exceed 79\% utilization.
\item \textbf{Bottleneck effect on global throughput:} In heterogeneous architectures, overall measured bandwidth does not scale linearly with the number of fast drives. Instead, global throughput is anchored to the baseline of the pure slow‑drive array (“4A”). This indicates that under traditional RAID scheduling semantics, the I/O processing rate of fast drives is forced down to match that of slow drives. Low‑level monitoring data collected by \texttt{iostat} further confirms this: for example, in the “3A‑1B” array under 4KB random reads, type‑A slow drives average 1780 MB/s, while the type‑B fast drive is similarly constrained to 1785 MB/s.
\end{itemize}

\subsubsection{Cause Analysis}

\begin{figure}[htb]
    \centering
    \subfloat[Homogeneous RAID]{
        \label{fig:c4:motiv:sbmho}
        \includegraphics[width=0.45\columnwidth]{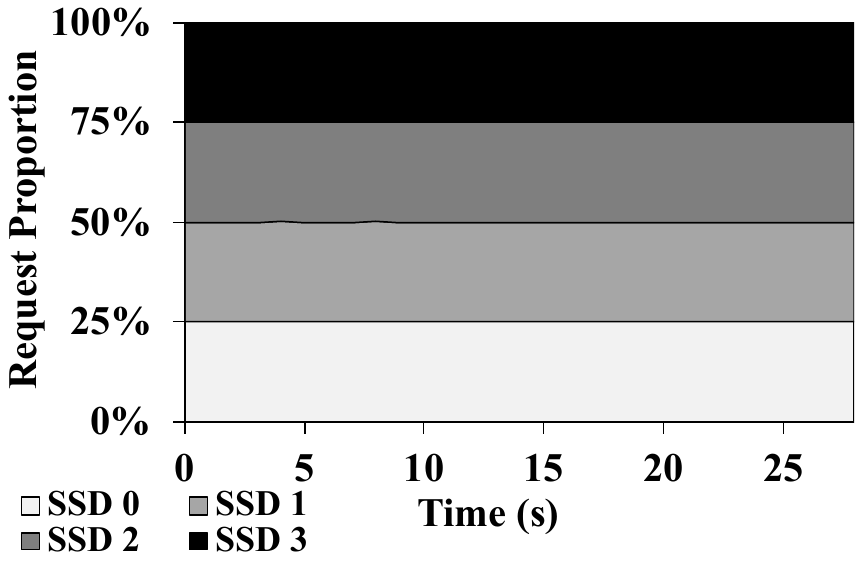}
    }
    \subfloat[Heterogeneous RAID]{
        \label{fig:c4:motiv:sbmhe}
        \includegraphics[width=0.45\columnwidth]{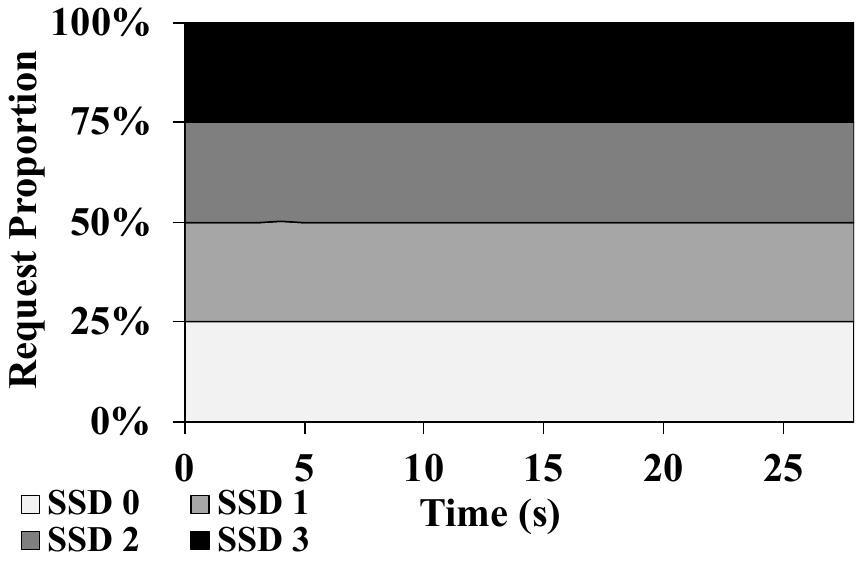} 
    }
    \caption{Proportion of requests queued at each SSD under 4KB random read workload}
 \label{fig:c4:motiv:sbm}
\end{figure}

\begin{figure}[htb]
    \centering
    \subfloat[Homogeneous RAID]{
        \label{fig:c4:motiv:qdho}
        \includegraphics[width=0.45\columnwidth]{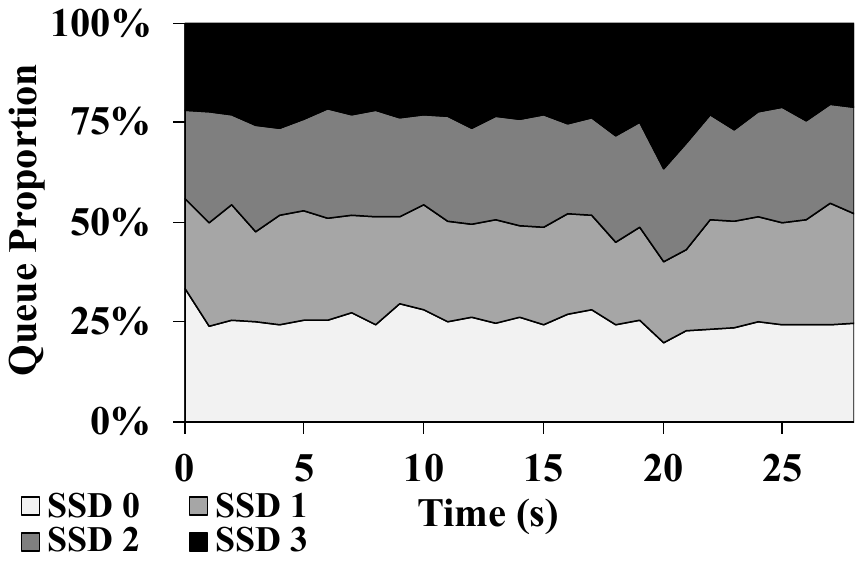}
    }
    \subfloat[Heterogeneous RAID]{
        \label{fig:c4:motiv:qdhe}
        \includegraphics[width=0.45\columnwidth]{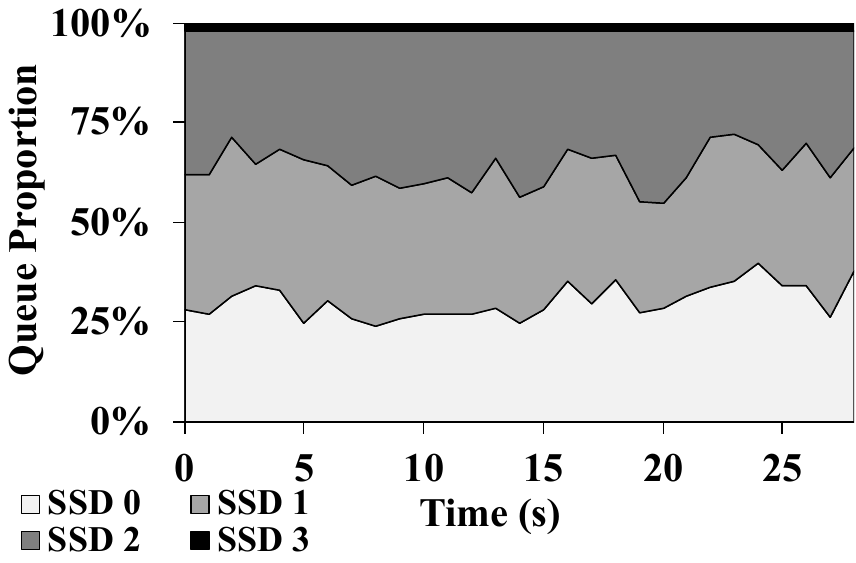} 
    }
    \caption{Proportion of requests handled by each SSD in the queue under 4KB random read workload}
 \label{fig:c4:motiv:qd}
\end{figure}

\begin{figure}[htb]
    \centering
    \subfloat[Homogeneous RAID]{
        \label{fig:c4:motiv:bwho}
        \includegraphics[width=0.45\columnwidth]{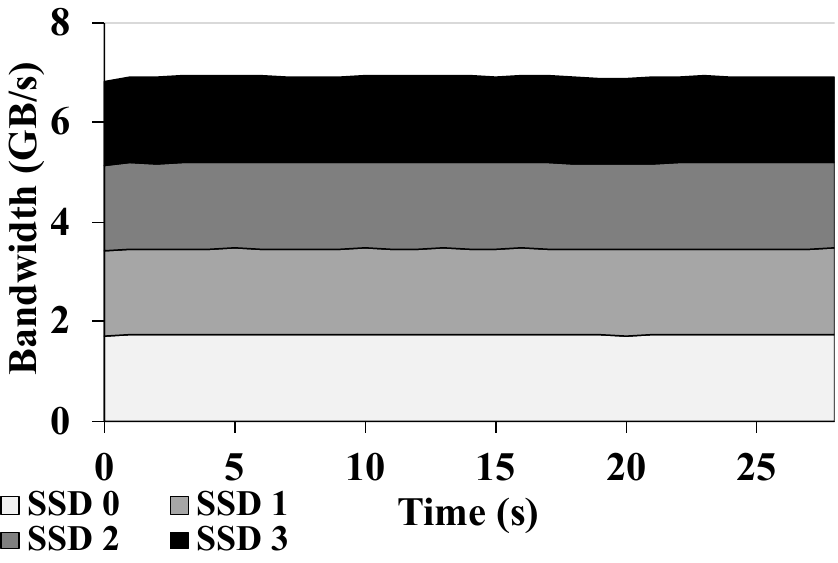}
    }
    \subfloat[Heterogeneous RAID]{
        \label{fig:c4:motiv:bwhe}
        \includegraphics[width=0.45\columnwidth]{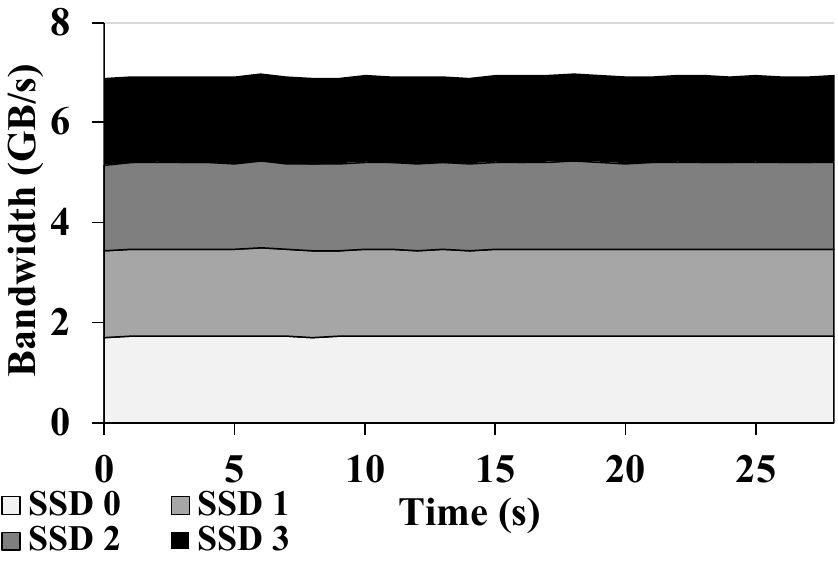} 
    }
    \caption{Bandwidth of each SSD under 4KB random read workload}
 \label{fig:c4:motiv:bwtime}
\end{figure}

To investigate the micro‑mechanisms behind bandwidth underutilization in heterogeneous arrays, we conducted fine‑grained analysis under 4KB random read workloads, examining request dispatch patterns and execution states of each member drive. At this granularity, read requests are aligned with stripe boundaries and never split across drives, meaning each I/O request is independently processed by a single backend SSD.

Two core metrics were defined: (i) \emph{queue depth proportion}, reflecting system concurrency, measured by high‑frequency sampling of RAID dispatch queues to calculate the share of in‑flight requests per SSD; and (ii) \emph{request arrival rate}, reflecting application‑side I/O pressure, measured as the proportion of initial requests routed to each physical drive per second. In addition, real‑time throughput of each drive was recorded. Homogeneous (“4A”) and heterogeneous (“3A‑1B”) arrays were used as baselines, with cumulative area plots shown in Figures~\ref{fig:c4:motiv:sbm}, \ref{fig:c4:motiv:qd}, and \ref{fig:c4:motiv:bwtime}.

Cross‑analysis of monitoring data yields the following insights:

First, striping semantics enforce absolute uniformity in request arrivals. As shown in Figures~\ref{fig:c4:motiv:sbmho} and \ref{fig:c4:motiv:sbmhe}, regardless of homogeneity or heterogeneity, the proportion of requests routed to each SSD remains around 25\% with minimal fluctuation. This indicates that simply introducing high‑performance drives does not alter the distribution of incoming requests.

Second, slow drives cause severe queue congestion. Comparing Figures~\ref{fig:c4:motiv:qdho} and \ref{fig:c4:motiv:qdhe}, homogeneous arrays show stable and balanced queue proportions (24.4\%–25.4\%, std.\ dev.\ $\leq$3.2\%). In contrast, heterogeneous arrays exhibit extreme divergence: slow drives’ queue proportions rise to 30.2\%–34.3\% (std.\ dev.\ $\leq$5\%), while fast drives drop to only 2.12\% (std.\ dev.\ $\leq$0.01\%). This reveals an I/O blocking propagation mechanism: slow drives accumulate unfinished requests, exhausting global queue resources and preventing new requests from being dispatched to idle fast drives. In production workloads, synchronous dependencies among read requests further exacerbate this effect, as lagging requests on slow drives block front‑end threads and hinder dispatch to fast drives.

Finally, high execution speed masks queue depth differences. As shown in Figures~\ref{fig:c4:motiv:qdhe} and \ref{fig:c4:motiv:bwhe}, although fast drives maintain extremely low queue depth, all drives exhibit nearly identical steady‑state throughput (mean std.\ dev.\ only 0.000158 MB/s). This demonstrates that low queue depth on fast drives is not due to fewer assigned requests, but rather their rapid completion of the 25\% share, leaving them idle for most observation periods.

In summary, the root cause of bandwidth underutilization in heterogeneous arrays lies in two conflicting perspectives:
\begin{enumerate}
    \item From the device side, fast SSDs require far higher request arrival rates to saturate their bandwidth.
    \item From the host side, constrained by fixed RAID striping, limited global queue depth, and application‑level dependencies, the system enforces nearly equal request distribution. Slow drives throttle the global dispatch rate, leaving fast drives idle and ultimately aligning all SSD bandwidth to that of the slowest drive.
\end{enumerate}

\subsection{Opportunities and Challenges of Caching Techniques}
\label{sec:c4:motiv:oppur}

The core solution to bandwidth underutilization lies in breaking the constraint of uniform request distribution and dynamically adjusting the dispatch rate to each member drive. While migrating hot data from slow SSDs to fast SSDs is possible, its adaptability is limited. In this work, we consider leveraging SSD caching for traffic diversion. Caching essentially redirects part of the request stream to cache devices, thereby regulating backend traffic. In theory, different diversion ratios can be applied to different SSDs, enabling differentiated request dispatch rates without affecting upper‑layer applications. Compared with DRAM caches, SSD caches provide larger capacity, and compared with direct data migration, they offer greater flexibility and safety. Furthermore, caching can improve overall read performance of the array. Traditional caching adopts a “serve‑on‑hit” model, where hit rate equals the proportion of traffic diverted to the cache, but lacks dynamic adjustment capability. Prior work such as NHC has noted that excessive diversion to SSD caches can be problematic~\cite{nhc,most}, and proposed dynamic adjustment to maximize combined cache and backend performance.

However, existing dynamic cache offloading mechanisms generally assume homogeneous backends and do not account for performance heterogeneity. For example, NHC introduces a global offloading probability scalar $p$: when a request hits in the cache, the system serves it directly with probability $p$, otherwise it is forwarded to the backend. Thus, the actual diversion ratio $\rho$ equals the product of physical hit rate $h$ and offloading probability $p$ ($\rho = h \cdot p$). Since optimal $p$ cannot be statically modeled under complex runtime conditions, online local search is required. NHC models a system‑level performance metric (e.g., aggregate bandwidth or latency) as an objective function $f(p)$. At iteration $t$, the system probes $f(p_t-s)$ and $f(p_t+s)$ with step size $s$, compares $f(p_t-s)$, $f(p_t)$, and $f(p_t+s)$, and selects the maximizing solution as $p_{t+1}$. Each probe requires sustained I/O sampling to obtain stable feedback.

\begin{figure}[htb]
    \centering
    \subfloat[Existing Cache Scheme]{
        \label{fig:c4:motiv:conceptnhc}
        \includegraphics[width=0.45\columnwidth]{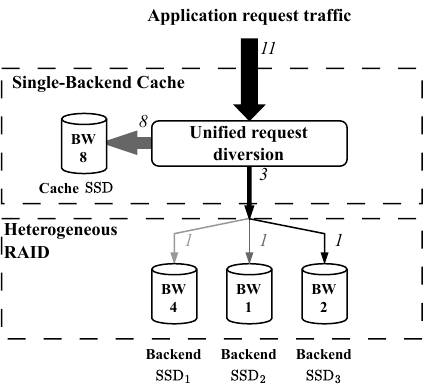}
    }
    \subfloat[HACache Scheme]{
        \label{fig:c4:motiv:conceptha}
        \includegraphics[width=0.45\columnwidth]{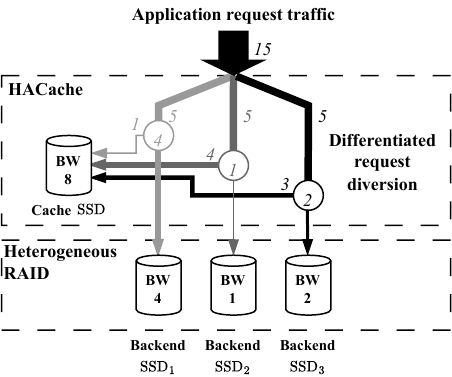} 
    }
    \caption{Comparison of cache schemes for heterogeneous RAID}
 \label{fig:c4:motiv:concept}
\end{figure}

While this one‑dimensional search converges efficiently in homogeneous arrays, it fails in heterogeneous scenarios. As shown in Figure~\ref{fig:c4:motiv:conceptnhc}, a uniform $p$ applies identical traffic reduction to all SSDs, leaving request arrival rates evenly distributed and failing to resolve queue blocking caused by slow drives. Empirical data confirms this: in a “3A‑1B” heterogeneous array with an additional SSDC cache under 4KB random reads, single‑$p$ optimization achieves only 13.79 GB/s aggregate bandwidth. This is merely the physical sum of the heterogeneous array and cache throughput, far below the theoretical peak of 18.31 GB/s, proving that fast drive B’s potential remains underutilized.

A possible remedy is to expand the parameter space, replacing scalar $p$ with a per‑drive vector $\mathbf{P}=\{p^j \mid 1 \le j \le N_{\text{SSD}}\}$, where $N_{\text{SSD}}$ is the number of member drives. However, this dramatically enlarges the search space, leading to dimensional explosion. For a 4‑drive array, the search branches per iteration increase from 3 to $3^4$. In general, as array size grows, the optimization space expands exponentially as $3^{N_{\text{SSD}}}$. Since each candidate state requires real device measurements over millisecond‑scale windows, such exponential online overhead is infeasible in enterprise RAID deployments with dozens of SSDs.

\section{HACache: Adaptive Cache Design for Heterogeneous RAID}
\subsection{Overview}

To address the problem of bandwidth underutilization in heterogeneous arrays, we propose an adaptive cache design for heterogeneous RAID, named \textit{HACache}. As illustrated in Figure~\ref{fig:c4:motiv:conceptha}, HACache introduces an SSD cache into the heterogeneous array to differentiate request diversion across backend SSDs. The proportion of traffic directed to each backend SSD is aligned with its bandwidth capacity, thereby maximizing utilization of all SSDs.

The core of this design lies in differentiated request diversion. To achieve this, three key challenges must be solved:
\begin{enumerate}
    \item How to determine the optimal diversion ratio for each backend SSD that maximizes aggregate bandwidth;
    \item How to approximate the above objective under complex runtime conditions;
    \item How to adjust the cache space allocation among backend SSDs when cache hit rate is limited, so that runtime behavior can approach the ideal diversion ratio as closely as possible.
\end{enumerate}

For the first challenge, this chapter provides a theoretical analysis to define the problem and establish the optimization objective. For the second challenge, HACache adopts a dynamic diversion ratio optimization strategy, which gradually approaches the target through a two‑phase adjustment mechanism at runtime. For the third challenge, HACache introduces a cache capacity regulation strategy, which adjusts cache allocation based on current hit capability and the optimized diversion ratio obtained in the second step.

\subsection{Optimization Objective of Cache Diversion Ratios}
\label{sec:c4:design:theo}

This section provides a formal analysis to compute the optimal diversion ratios from a theoretical perspective.

\subsubsection{Formal Problem Definition}

We first formalize the optimization problem of cache diversion ratios in heterogeneous arrays. Consider a heterogeneous backend array consisting of $N_{\text{SSD}}$ physical drives and a dedicated SSD cache layer. Let the maximum physical bandwidth of the $i$‑th backend SSD ($i \in [1, N_{\text{SSD}}]$) be $b_{max,i}$, and the maximum physical bandwidth of the cache SSD be $c_{max}$.

During any steady‑state period of system operation, we define the following state variables:
\begin{itemize}
    \item $b_i$: the actual physical bandwidth sustained by the $i$‑th backend SSD.
    \item $c_i$: the effective bandwidth component served by the cache SSD on behalf of requests originally routed to the $i$‑th backend SSD.
    \item $\rho_i$: the diversion ratio for the $i$‑th backend SSD, i.e., the proportion of its logical traffic ($b_i+c_i$) served by the cache, satisfying $c_i = \rho_i(b_i+c_i)$.
\end{itemize}

To construct a solvable theoretical model, we establish the following system‑level assumptions:
\begin{enumerate}
    \item \textbf{Load saturation:} The front‑end application issues sufficiently deep I/O queues such that the system is always fully loaded. Thus, among feasible combinations of $b_i$ and $c_i$, the system seeks the maximum values.
    \item \textbf{Uniform logical distribution:} Based on the striping property revealed in Section~\ref{sec:c4:motiv:phenomenon}, logical traffic routed to each member drive is equal. Therefore, $\forall i_1, i_2 \in [1, N_{\text{SSD}}], b_{i_1} + c_{i_1} = b_{i_2} + c_{i_2}$.
    \item \textbf{Known performance boundaries:} The peak bandwidths of all physical devices ($b_{max,i}$ and $c_{max}$) are known and stable.
    \item \textbf{Ideal cache diversion capability:} Cache space is assumed to be abundant, supporting a physical hit rate of 100\%. Under this extreme condition, the cache can sustain any diversion ratio $\rho_i \in [0,1]$, allowing us to derive the theoretical upper bound of system performance.
\end{enumerate}

Based on these definitions and assumptions, the task of determining the optimal diversion ratios for backend SSDs is transformed into a constrained multivariable optimization problem: under hardware throughput limits, find the optimal set $\{\rho_i \mid i \in [1, N_{\text{SSD}}]\}$ that maximizes global effective read bandwidth. The complete mathematical formulation is as follows:

\begin{align}
\max_{\{\rho_i \mid i \in [1, N_{\text{SSD}}]\}} \quad & \sum_{i=1}^{N_{\text{SSD}}} (b_i + c_i) \label{eq:c4:motiv:objective} \\
\text{s.t.} \quad 
& c_i = \rho_i(b_i + c_i), \quad \forall i \in [1, N_{\text{SSD}}] \label{eq:c4:motiv:constraint_a} \\
& b_{i_1} + c_{i_1} = b_{i_2} + c_{i_2}, \quad \forall i_1, i_2 \in [1, N_{\text{SSD}}] \label{eq:c4:motiv:constraint_b} \\
& 0 \le \sum_{i=1}^{N_{\text{SSD}}} c_i \le c_{max} \label{eq:c4:motiv:constraint_c} \\
& 0 \le b_i \le b_{max,i}, \quad \forall i \in [1, N_{\text{SSD}}] \label{eq:c4:motiv:constraint_d}
\end{align}

\subsubsection{Problem Solving}

The above constrained optimization problem can be heuristically solved using the classical \textbf{water‑filling algorithm} idea. Based on this, we design the \textbf{Optimal Cache Diversion Ratio Planning Algorithm}, whose core logic is to allocate the limited global cache bandwidth resource in a coordinated manner, thereby compensating bottlenecks and maximizing overall array throughput.

We introduce an auxiliary variable $T$ to represent the logical total bandwidth of each member drive (i.e., $T = b_i + c_i$). According to the “uniform logical distribution” constraint described earlier, the original objective of maximizing global bandwidth can be equivalently transformed into finding the optimal logical bandwidth baseline (the “water level”) $T^*$, such that the global aggregate bandwidth $N_{\text{SSD}} \cdot T^*$ is maximized under physical device limits.

The complete derivation and iterative process are shown in Algorithm~\ref{alg:c4:design:optimal}. The algorithm first injects limited cache bandwidth into the most constrained slow drives. By gradually expanding the set of backend SSDs covered by the cache, it fills performance gaps bottom‑up and iteratively approaches the global optimal logical water level $T^*$ (lines 3–10). Once convergence anchors the optimal baseline $T^*$, the algorithm combines each backend device’s physical bandwidth limit to precisely back‑calculate the ideal diversion ratio for each member drive (lines 11–13).

\begin{algorithm}[htb]
    \caption{Optimal Cache Diversion Ratio Algorithm}
    \label{alg:c4:design:optimal}
    \begin{algorithmic}[1]
        \REQUIRE $\{b_{max,i}|i\in [1,N_{\text{SSD}}]\}$, maximum bandwidth limits of SSDs;
        \REQUIRE $c_{max}$, cache bandwidth;
        \ENSURE $\{\rho_i\}$, diversion ratios for each SSD;
        
    \COMMENT Sort backend SSD bandwidth limits in ascending order;
    \STATE Sort $\{b_{\max,i}\}$ as $b_{(1)} \le b_{(2)} \le \dots \le b_{(N)}$;
    
    \COMMENT Initialize cumulative sum of backend bandwidth;
    \STATE $SumB \gets 0$;
    
    \COMMENT Iterate to find optimal $k$ and $T^*$;
    \FOR{$k \gets 1$ \TO $N_{\text{SSD}}$}
        \STATE $SumB \gets SumB + b_{(k)}$;
        \STATE $T_{temp} \gets (c_{\max} + SumB) / k$;
        \IF  {$k=N_{\text{SSD}}$ \OR $T_{temp} \le b_{(k+1)}$}
            \STATE $T^* \gets T_{temp}$;
            \STATE \textbf{break};
        \ENDIF
    \ENDFOR
    
    \COMMENT Compute optimal diversion ratios based on $T^*$;
    \FOR{$i \gets 1$ \TO $N_{\text{SSD}}$}
        \STATE $\rho_i \gets \max\left(0, 1 - \frac{b_{\max,i}}{T^*}\right)$;
    \ENDFOR
\end{algorithmic}
\end{algorithm}

\begin{figure}[htb]
    \centering
    \subfloat[$c_{max}=4,\{b_{max}\}=\{2,3,3,5\}$]{
        \label{fig:c4:design:optexa}
        \includegraphics[width=.95\linewidth]{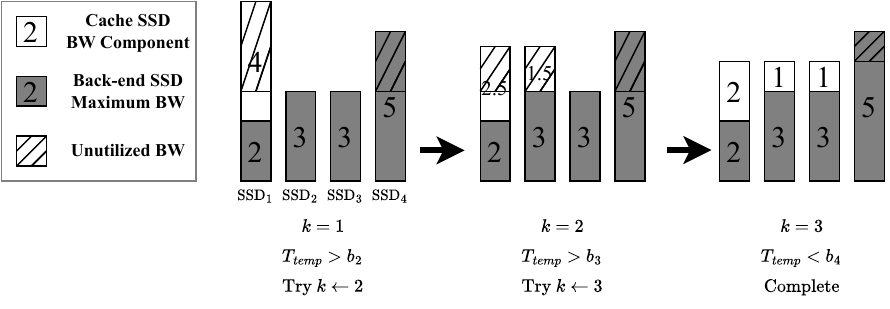}
    }
    \\
    \subfloat[$c_{max}=11,\{b_{max}\}=\{2,3,3,5\}$]{
        \label{fig:c4:design:optexb}
        \includegraphics[width=.95\linewidth]{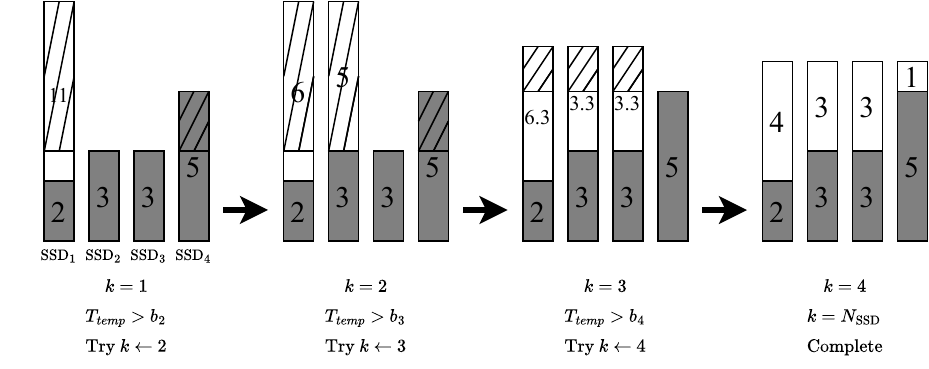} 
    }
    \caption{Examples of the Optimal Cache Diversion Ratio Algorithm}
 \label{fig:c4:design:optex}
\end{figure}

We illustrate the rationality of the optimal cache diversion ratio algorithm with two examples in Figure~\ref{fig:c4:design:optex}, showing different diversion outcomes. The RAID consists of four SSDs (SSD1–SSD4). In Figure~\ref{fig:c4:design:optexa}, we demonstrate a case where cache cannot fully resolve bandwidth underutilization. The cache SSD has a maximum bandwidth of 4, while backend SSDs have maximum bandwidths of 2, 3, 3, and 5. The algorithm proceeds as follows:
\begin{itemize}
    \item \textbf{First attempt (covering SSD1):} Allocate all cache bandwidth to the slowest SSD1. Its logical bandwidth $T_{temp}$ rises to 6 (2+4), yielding \{6,3,3,5\}. Since the global bottleneck shifts to SSD2 (3), excess cache for SSD1 is wasted, prompting expansion.
    \item \textbf{Second attempt (covering SSD1–2):} Distribute cache evenly to SSD1 and SSD2. Their combined bandwidth is 5, plus cache 4, giving $T_{temp}=4.5$. The logical set becomes \{4.5,4.5,3,5\}. SSD3 (3) becomes the bottleneck, requiring further expansion.
    \item \textbf{Third attempt (covering SSD1–3):} Pool cache across SSD1–3. Their sum is 8, plus cache 4, yielding $T_{temp}=4$. The logical set becomes \{4,4,4,5\}. Cache bandwidth is fully utilized, and the system reaches the theoretical maximum under hardware constraints. The algorithm terminates and computes diversion ratios accordingly.
\end{itemize}

In Figure~\ref{fig:c4:design:optexb}, we show a case where cache fully resolves bandwidth underutilization. The first steps are similar, until cache is distributed across all four backend SSDs.

\subsection{Dynamic Adjustment of Cache Diversion Ratios}
\subsubsection{Limitations of the Optimal Cache Diversion Ratio Algorithm}

The optimal cache diversion ratio algorithm proposed in Section~\ref{sec:c4:design:theo} can compute diversion ratios for backend SSDs under relatively stable and deterministic workloads. However, its applicability in complex runtime environments remains limited. Specifically, while assumptions regarding request quantity and distribution are reasonable and easily satisfied, assumptions about cache hit rate and bandwidth limits may not hold, affecting the accuracy of the optimal algorithm.

Regarding the hit rate assumption, cache hit rate may not support the optimal diversion ratio depending on workload locality and cache size, and runtime hit rate may vary. A fixed diversion ratio may therefore be suboptimal.

\begin{figure}[htb]
  \centering
  \includegraphics[width=.8\linewidth]{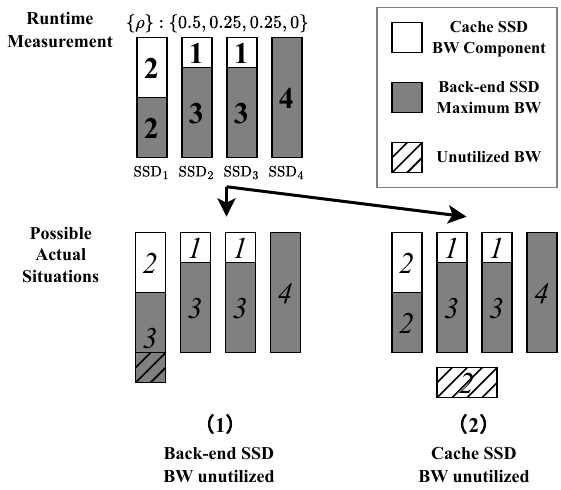}
  \caption{Different possible device bandwidth limits corresponding to the same SSD measurement values}
\label{fig:c4:design:measurement} 
\end{figure}

Regarding the bandwidth limit assumption, SSD bandwidth limits are difficult to specify. First, SSD bandwidth varies significantly with I/O granularity, and different SSDs exhibit different trends. For example, as shown in Table~\ref{tab:c4:eval:ssds}, SSD~A achieves 1800 MB/s under 4KB random reads, which is only about 51\% of its 3500 MB/s bandwidth under 128KB random reads. Second, real workloads often involve mixed granularities, making offline profiling complex. Finally, runtime systems cannot directly determine SSD bandwidth limits under current workloads, since it is unclear whether drives have idle capacity. When initial diversion ratios are inaccurate or workload characteristics change, ratios may become inappropriate. Due to constraint (4.3), both backend SSDs and cache SSDs may have unused bandwidth under such conditions. Figure~\ref{fig:c4:design:measurement} illustrates an example RAID with four SSDs (SSD1–SSD4), where $\{c_i|i \in[1,4]\}=\{2,1,1,0\}$. Two possible cases arise under the same measurement values:

\begin{enumerate}
    \item Case (1) shows unused backend SSD bandwidth. Since other SSDs and cache sum to 4, constraint (4.3) forces SSD1 and cache to total 4. With SSD1’s diversion ratio at 0.5, its exposed bandwidth is $4 \times 0.5 = 2$, leaving 1 unused. Ideally, cache diversion should be $\{1.25,1.25,1.25,0.25\}$.
    \item Case (2) shows unused cache bandwidth. Due to inappropriate diversion ratios, after serving specified proportions, the cache still has 2 bandwidth units left. Ideally, cache diversion should be $\{2.5,1.5,1.5,0.5\}$.
\end{enumerate}

In both cases, runtime bandwidth does not reflect SSD maximum capacity. Moreover, depending on diversion ratios and SSD bandwidths, both situations may occur simultaneously across multiple SSDs, further complicating runtime determination of maximum bandwidth. Therefore, an algorithm that does not rely on predefined bandwidth limits is necessary.

\subsubsection{Two-Phase Dynamic Valve Value Adjustment Strategy}

We aim to design an algorithm that can automatically search for optimized diversion ratios at runtime based solely on observed bandwidth, and that can converge from any initial diversion state toward the optimal state.

To address the limitation of cache hit rate, the algorithm regulates the diversion probability $p_i$ after a cache hit, instead of directly optimizing the overall diversion ratio $\rho_i$. Specifically, as discussed in Section~\ref{sec:c4:motiv:oppur}, in dynamic diversion design we have $\rho_i = h_i p_i$, where $h_i$ is the cache hit rate for requests associated with backend SSD $i$. Since $p_i$ controls the diversion ratio of cache‑hit requests for SSD $i$, we refer to it as the \textit{valve value}. In addition, HACache introduces cache capacity regulation to adjust cache allocation among backend SSDs, thereby optimizing hit rate under limited cache space (details in Section~\ref{sec:c4:design:cap}).

To address uncertainty in bandwidth limits, our approach divides the algorithm into two phases: the first phase reduces idle bandwidth of backend SSDs, and the second phase reduces idle bandwidth of the cache SSD. These two phases alternate until the optimal diversion ratio is reached. Based on this idea, HACache proposes a two‑phase dynamic diversion adjustment strategy, as shown in Algorithm~\ref{alg:c4:design:2phase}.

\begin{algorithm}[!htbp]
    \caption{Two-Phase Dynamic Valve Value Adjustment Strategy}
    \label{alg:c4:design:2phase}
     \begin{algorithmic}[0]
    \STATE \textbf{$p_{i}$:} valve value of backend $\text{SSD}_i$
    \STATE \textbf{$P$:} configuration set, i.e., $\{p_{i}|i\in[1,N_\textbf{SSDs}]\}$
    \STATE \textbf{$h_i$:} cache hit rate for requests associated with $\text{SSD}_i$
    \STATE \textbf{$b_i(P)$:} bandwidth of $\text{SSD}_i$ under configuration $P$
    \STATE \textbf{$T(P)$:} sum of backend SSD bandwidth and corresponding cache bandwidth under $P$
    \STATE \textbf{$S(P)$:} system aggregate bandwidth under $P$, i.e., RAID bandwidth plus cache SSD bandwidth
    \STATE \textbf{$Pmod(P,p_i\gets p_i')$:} configuration set after updating $p_i$ to $p_i'$
    \STATE \textbf{$ExpV(T,b,h)$:} valve value estimation algorithm
    \end{algorithmic}
    \begin{algorithmic}[1]
        
        \WHILE{true}
            \STATE  // Phase 1: replace cache bandwidth with idle backend SSD bandwidth
            \FOR{$i \gets 1$ \TO $N_{\text{SSD}}$}
                \WHILE{true}
                    \STATE Select expected bandwidth increment $\Delta b_i$ for $\text{SSD}_i$
                    \STATE $P'\gets Pmod(P, p_i \gets ExpV(T(P),b_i+\Delta b_i, h_i))$
                    \IF{$T(P') < T(P)$ \OR $P'=P$}
                        \STATE $P\gets Pmod(P, p_i \gets ExpV(T(P),b_i(P'), h_i))$
                        \STATE \textbf{break}
                    \ELSE
                        \STATE $P\gets P'$
                    \ENDIF
                \ENDWHILE
            \ENDFOR
            
            \STATE // Phase 2: evenly distribute idle cache bandwidth across all backend SSDs
            \WHILE{true}
                \STATE Select cache bandwidth increment $\Delta c$
                \STATE $P'\gets P$
                \FOR{$i \gets 1$ \TO $N_{\text{SSD}}$}
                    \STATE $T_{Exp} \gets T(P)+\Delta c/N_\text{SSDs}$
                    \STATE $P'\gets Pmod(P’,p_i \gets ExpV(T_{Exp},b_i(P), h_i)$
                \ENDFOR
                \IF{$S(P') < S(P)$ \OR $P'=P$}
                    \STATE \textbf{break}
                \ELSE
                    \STATE $P\gets P'$
                \ENDIF
            \ENDWHILE
        \ENDWHILE
    \end{algorithmic}
\end{algorithm}

\begin{algorithm}[htb]
    \caption{Valve Value Estimation Algorithm}
    \label{alg:c4:design:exp}
     \begin{algorithmic}[1]
    \REQUIRE $h$: cache hit rate for SSD
    \REQUIRE $b$: expected bandwidth of SSD
    \REQUIRE $T$: expected logical bandwidth of SSD including cache component
    \ENSURE $p$: expected valve value
    \STATE $\rho \gets 1-b/T$
    \STATE $p \gets \rho/h$
    
    \COMMENT Restrict $p$ to $[0,1]$
    \STATE $p \gets min(p,1)$
    \STATE $p \gets max(p,0)$
     \end{algorithmic}
\end{algorithm}

In Phase 1, HACache attempts to replace cache bandwidth with idle backend SSD bandwidth. Line 4 selects increments according to implementation strategy (e.g., fixed values or gradually decreasing during convergence). Line 5 computes new valve values based on expected bandwidth, first back‑calculating from diversion ratio, then restricting $p$ to [0,1] to avoid invalid ratios. If $\text{SSD}_i$ has sufficient idle bandwidth, $T_i$ remains unchanged (with $b_i$ increasing and $c_i$ decreasing), and HACache continues adjustment. Otherwise, $b_i$ remains unchanged, diversion ratio shrinks, $c_i$ decreases, and $T_i$ decreases. At this point, rollback is required and adjustment ends for $\text{SSD}_i$. If valve limits cause no modification, the loop also terminates.

After Phase 1 maximizes backend SSD bandwidth, Phase 2 redistributes reclaimed or unused bandwidth evenly across backend SSDs. Due to constraint (4.3), valve adjustments in Phase 2 must be synchronized across all SSDs. If cache SSD has sufficient bandwidth, system bandwidth increases; otherwise, overall bandwidth decreases, requiring rollback and termination of Phase 2.

The two phases repeat alternately. Algorithm~\ref{alg:c4:design:2phase} does not specify termination conditions, allowing continuous adjustment under changing workloads. For stable workloads, if either phase produces no modification to $P$, the strategy terminates, indicating convergence.

\begin{figure} [htb]
  \centering
  \includegraphics[width=.8\linewidth]{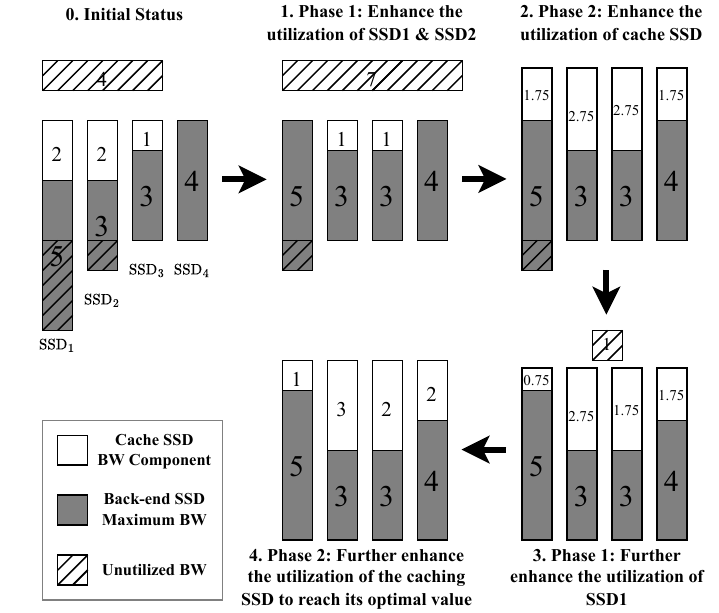}
  \caption{Example of Two-Phase Dynamic Valve Value Adjustment Strategy}
\label{fig:c4:design:2phaseexp} 
\end{figure}

Figure~\ref{fig:c4:design:2phaseexp} illustrates an example of the two‑phase dynamic valve value adjustment strategy. Initially, backend $\text{SSD}_1$, $\text{SSD}_2$, and the cache SSD have idle bandwidth. State 1 shows Phase 1 adjustments: HACache increases bandwidth of $\text{SSD}_1$ by 2 and $\text{SSD}_2$ by 1, but cannot increase $\text{SSD}_3$ or $\text{SSD}_4$. Due to other SSD limits, $\text{SSD}_1$ still has 1 unused bandwidth. State 2 shows Phase 2 redistribution of 7 unused bandwidth units evenly across four SSDs. With additional cache diversion, State 3 shows Phase 1 successfully utilizing the remaining 1 bandwidth of $\text{SSD}_1$, replacing 1 cache bandwidth. Finally, State 4 shows Phase 2 redistributing this 1 cache bandwidth evenly, achieving optimal state. Further iterations produce no new optimization.

\subsection{Cache Capacity Regulation}
\label{sec:c4:design:cap}

\begin{figure} [htb]
  \centering
  \includegraphics[width=.8\linewidth]{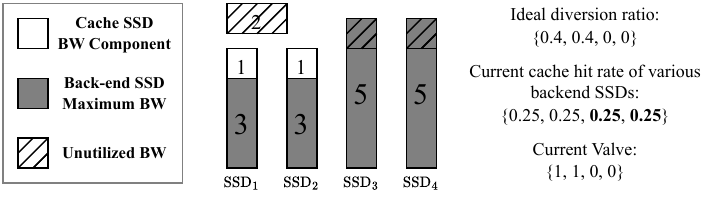}
  \caption{Example of cache capacity mismatch among backend SSDs}
\label{fig:c4:design:hrexp} 
\end{figure}

Considering $\rho_i = h_i p_i$, the cache hit rate represents the maximum achievable diversion ratio. Although HACache cannot increase the overall cache hit rate, it can optimize cache allocation by adjusting the space assigned to each backend SSD. On the cache device, each cached block corresponds uniquely to a backend SSD; we define the total capacity of cached blocks associated with a backend SSD as its \emph{cache capacity}. Figure~\ref{fig:c4:design:hrexp} illustrates a potential optimization opportunity. As shown, backend SSD bandwidth limits are $\{3,3,5,5\}$, and the cache SSD has a total bandwidth of 4. To fully utilize all bandwidth, the ideal diversion ratios are $\{0.4,0.4,0,0\}$, meaning SSD1 and SSD2 each receive 2 units of cache bandwidth, while SSD3 and SSD4 require none. Due to striping distribution, non‑differentiated cache admission often yields similar access patterns across SSDs, leading to convergent hit rates. Suppose each backend SSD has a hit rate of 0.25; then the maximum diversion ratio for each is limited to 0.25. For SSD3 and SSD4, even with valve values of 1, diversion ratios remain capped at 0.25, preventing full utilization of cache bandwidth. Conversely, SSD3 and SSD4 require no diversion (valve values of 0), so caching their data wastes space. Reallocating cache from SSD3 and SSD4 to SSD1 and SSD2 can improve overall system bandwidth.

HACache should sense each backend SSD’s demand for cache space and regulate capacity to achieve differentiated hit rate control. Demand can be inferred from valve values. In Figure~\ref{fig:c4:design:hrexp}, $p_1$ and $p_2$ equal 1, indicating insufficient cache capacity for SSD1 and SSD2, which may require higher hit rates to improve system bandwidth. Meanwhile, $p_3$ and $p_4$ are significantly below 1, indicating excess capacity. HACache uses a preset threshold $p_{thres}$ to determine excess capacity: if $p_i < p_{thres}$, SSD $i$ has surplus cache capacity and can be reduced; if $p_i = 1$, SSD $i$ lacks capacity and requires expansion. Not using $p_i < 1$ as the sole criterion allows HACache to retain some cache margin for each SSD, tolerating workload fluctuations and avoiding re‑filling when demand shifts.

For each adjustment, HACache uses a preset reclamation granularity $\Delta q$. For each SSD with surplus capacity, HACache checks whether reclaiming $\Delta q$ would cause insufficiency. For stack‑based replacement policies (e.g., LRU, LFU), miss‑rate curve estimation~\cite{mrc1,mrc2,mrc3} can predict post‑reclamation hit rate $h_i^*$; if $h_i^* < h_i p$, reclamation is skipped. For non‑stack policies (e.g., FIFO), HACache partitions cache into shards (with $\Delta q$ as shard size), each shard containing data from one SSD. HACache computes shard contribution $hit_{shard}/hit_{SSD_i}$; if $hit_{shard}/hit_{SSD_i}+p_i > 1$, reclamation is skipped. Reclaimed capacity is evenly redistributed to SSDs lacking cache. HACache avoids reclaiming from SSDs that previously received capacity, preventing oscillation and ensuring convergence.

\begin{figure} [htb]
  \centering
  \includegraphics[width=\linewidth]{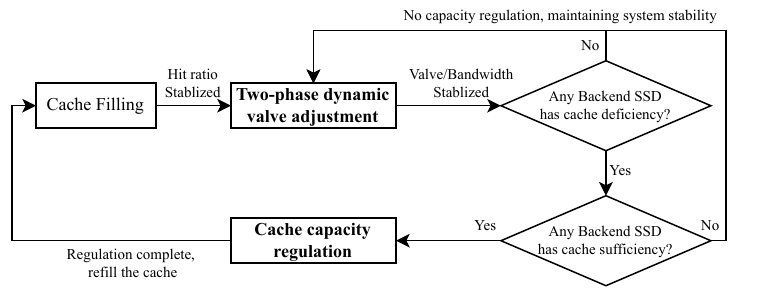}
  \caption{Complete execution flow of HACache}
\label{fig:c4:design:whexec} 
\end{figure}

Thus, the full optimization workflow of HACache is shown in Figure~\ref{fig:c4:design:whexec}. This process ensures that under dynamic workloads, the system continuously approaches local optima and ultimately achieves global resource optimization. After startup, the system enters cache warm‑up. As hot data is cached, HACache monitors hit rates. Once hit rates stabilize, indicating the working set is captured, valve value optimization begins. Through iterative two‑phase adjustment, valve values and aggregate bandwidth converge to a steady state, marking saturation under current capacity. The system then initiates cache capacity evaluation, consisting of two checks: first, whether any SSD lacks capacity; if none, the system maintains current allocation. If insufficiency exists, HACache checks for surplus capacity; if none, allocation remains unchanged. Only when both “demand” and “supply” exist does HACache trigger capacity regulation. Since transferred cache contents are invalidated, the system refills them, stabilizes hit rates, and re‑enters two‑phase valve adjustment, repeating the cycle.

\section{Implementation and Discussion}
\label{sec:c4:imple}

We implemented a prototype of HACache based on SPDK~\cite{spdk} as a proof of concept.  
The implementation consists of approximately 3000 lines of code. HACache does not impose restrictions on the cache management algorithm; for simplicity, our prototype adopts a FIFO‑like policy. Regarding the choice of $\Delta b_i$ and $\Delta c$ in the two‑phase dynamic valve value adjustment strategy, larger values accelerate convergence toward the target configuration set, but may cause performance fluctuations when rollback is required (Algorithm~\ref{alg:c4:design:2phase}, lines 7 and 21). Our current implementation uses fixed values: $\Delta b_i$ is set to 1 GB, and $\Delta c$ is set to $100\text{MB}\times N_{\text{SSDs}}$ to avoid impractically small bandwidth adjustments. For cache capacity regulation, we divide the cache space into 256 shards, set $\Delta q$ to 8 shards, and choose $p_{thres}=0.9$.

\begin{figure}[htb]
    \centering
    \subfloat[4KB Saturation Latency]{
        \label{fig:c4:imple:lb4}
        \includegraphics[width=.22\columnwidth]{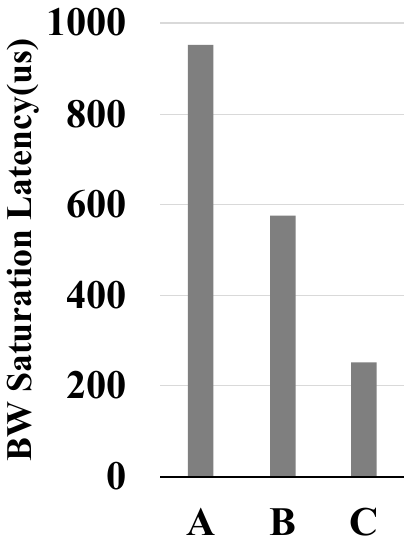}
    }
    \subfloat[128KB Saturation Latency]{
        \label{fig:c4:imple:lb8}
        \includegraphics[width=.22\columnwidth]{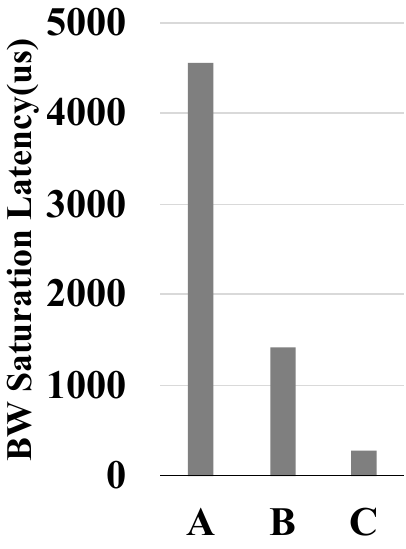} 
    }
    \subfloat[4KB Saturation Queue Depth]{
        \label{fig:c4:imple:db4}
        \includegraphics[width=.22\columnwidth]{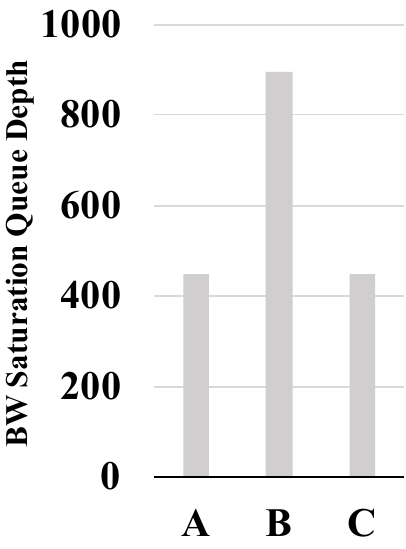}
    }
    \subfloat[128KB Saturation Queue Depth]{
        \label{fig:c4:imple:db8}
        \includegraphics[width=.22\columnwidth]{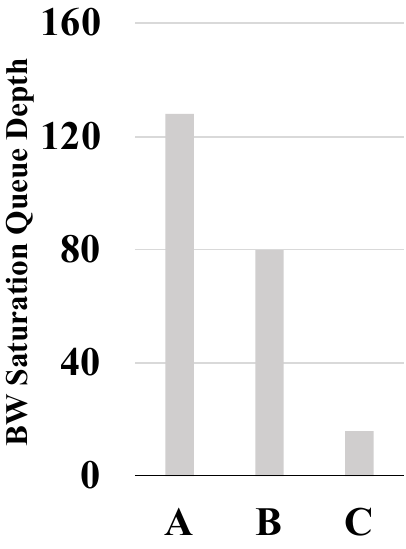} 
    }
    \caption{Saturation latency and cumulative queue depth of different SSDs under 4KB and 128KB random read workloads}
 \label{fig:c4:imple:satur}
\end{figure}

HACache directly uses bandwidth as the monitoring metric to determine whether SSDs are saturated. An alternative approach could use latency or queue depth, but HACache optimizes bandwidth itself, and these parameters do not directly correlate with bandwidth. Specifically, the relationship between bandwidth and latency or queue depth varies significantly across SSD models and workload characteristics. We measured SSD A, SSD B, and SSD C under 4KB and 128KB random read workloads using fio, with varying thread counts and queue depths. We defined the saturation point as the bandwidth at 99\% of the maximum, and recorded the corresponding I/O latency and cumulative queue depth (threads $\times$ queue depth) as saturation latency and saturation queue depth. Results are shown in Figure~\ref{fig:c4:imple:satur}. Across the four subfigures, we observe significant differences in saturation latency and queue depth among SSDs under the same workload. Even SSD B and SSD C, which have similar performance, exhibit notable differences in saturation latency and queue depth. Furthermore, modeling saturation for each SSD is difficult, as Figures~\ref{fig:c4:imple:lb4} and \ref{fig:c4:imple:lb8}, as well as Figures~\ref{fig:c4:imple:db4} and \ref{fig:c4:imple:db8}, show that even the same SSD exhibits different saturation latency and queue depth under different I/O granularities. Therefore, HACache ultimately uses bandwidth directly to determine saturation.

\section{Evaluation}
\subsection{Experimental Setup}
\label{sec:c4:eval:setup}

\subsubsection{Experimental Platform}
We use a Lenovo ST650 V3 server equipped with two 16‑core Intel Xeon Gold 5416S processors and 256 GB DDR5 memory, running Ubuntu 24.04 with Linux kernel v6.14.0.  
The specifications of the three SSDs used are listed in Table~\ref{tab:c4:eval:ssds}. SSD A connects to the host via PCIe 3.0, while SSD B and SSD C both use PCIe 4.0.

\subsubsection{Comparison Schemes}
The RAID configuration adopts the RAID5F format in SPDK. Unlike classic RAID5, RAID5F only supports full‑stripe writes, but its read behavior is identical to RAID5. Since our focus is on read workloads, this difference does not affect our evaluation.  
When testing different RAID compositions, for simplicity we denote homogeneous RAIDs by the number of drives plus SSD type (e.g., “4A”, “3A‑1B”), consistent with Section~\ref{sec:c4:motiv:phenomenon}. Similar to that section, SSD A is treated as the slow SSD and SSD B as the fast SSD in RAID compositions. In addition, one SSD C is used as the cache SSD.  
For comparison, we also implement the non‑hierarchical cache scheduler (NHC), which treats the entire RAID as the capacity tier and the cache SSD as the performance tier, optimizing bandwidth as the target metric. We do not directly compare with RAID without cache, since it lacks the bandwidth of the cache SSD and its performance is guaranteed to be lower than NHC.  
We also report the aggregate bandwidth of all devices in each configuration (i.e., the sum of all RAID SSDs and the cache SSD), which serves as the upper bound of system bandwidth.

\subsubsection{Workload Setup}
We use \texttt{fio} as the workload generator, with a default concurrency of 16 threads and a queue depth of 64 per thread.

\begin{table}[!htbp]
 \centering
\small
  \caption{Specifications of SSDs used in evaluation} 
  \label{tab:c4:eval:ssds}
  \begin{tabular}{ccc}
    \toprule
    \textbf{SSD Model} & \textbf{4KB Random } & \textbf{128KB Sequential} \\
    \textbf{} &\textbf{Write Bandwidth} &\textbf{Write Bandwidth}\\
    \midrule
    \textbf{A} & 1800 MB/s   & 3500 MB/s\\
    \textbf{B} & 6350 MB/s  & 7100 MB/s\\
    \textbf{C} & 7000 MB/s & 7100 MB/s \\
    \bottomrule
\end{tabular}
\end{table}

\subsection{Evaluation under Sufficient Cache Capacity}

We first evaluate HACache in scenarios where cache capacity is sufficient.  
For the workload, we set cache size to 10\% of the I/O range and configure fio so that 5\% of the space accounts for 95\% of accesses. In this way, hot data can be fully cached, and the system ultimately achieves a 95\% hit rate. We test two workloads: 4KB random reads and 128KB random reads. In addition to heterogeneous RAID configurations “1A‑3B”, “2A‑2B”, and “3A‑1B”, we also test homogeneous RAIDs “4A” and “4B” to demonstrate HACache’s applicability to homogeneous arrays.

\begin{figure} [htb]
  \centering
  \includegraphics[width=.8\linewidth]{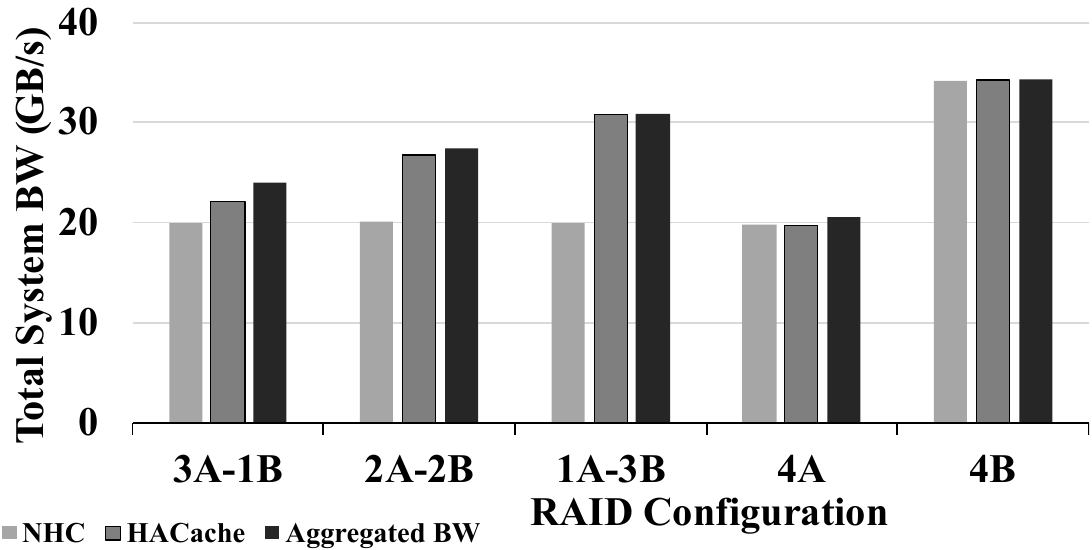}
  \caption{Performance comparison under sufficient cache capacity with 128KB random reads}
\label{fig:c4:eval:128kp} 
\end{figure}

Results for 128KB random reads are shown in Figure~\ref{fig:c4:eval:128kp}.  
For heterogeneous RAIDs, NHC achieves on average only 73.3\% of aggregate bandwidth, while HACache achieves 96.3\%. Moreover, the less efficiently heterogeneous RAIDs utilize bandwidth, the larger the gap between HACache and NHC. In the “3A‑1B” configuration, HACache outperforms NHC by 8\% aggregate bandwidth, whereas in “1A‑3B” the gap widens to 35.1\%. This is because HACache leverages cache to compensate for slow drives, thereby releasing otherwise unused bandwidth. In “3A‑1B”, cache SSD bandwidth is insufficient to fully bridge the gap between SSD A and SSD B, leaving some bandwidth of SSD B unused (similar to the scenario in Figure~\ref{fig:c4:design:optexa}). As a result, HACache achieves only 92\% of aggregate bandwidth, while the other two heterogeneous configurations reach at least 97\%. Even so, HACache still utilizes part of SSD B’s unused bandwidth, outperforming NHC. For homogeneous RAIDs, both “4A” and “4B” achieve bandwidth close to the aggregate limit, showing HACache’s adaptability even in homogeneous settings.

For 4KB random reads, the overall trend is similar. NHC achieves on average only 61.1\% of aggregate bandwidth, while HACache achieves 85.8\%. Homogeneous RAIDs again perform close to aggregate bandwidth. Since 4KB workloads require higher concurrency to fully exploit bandwidth, prototype overhead and platform concurrency settings result in slightly weaker performance. In “3A‑1B” and “2A‑2B”, SSD B’s bandwidth is not fully utilized, yielding about 84\% of aggregate bandwidth. In “1A‑3B”, HACache achieves 89.9\% of aggregate bandwidth.

\begin{figure} [htb]
  \centering
  \includegraphics[width=.8 \linewidth]{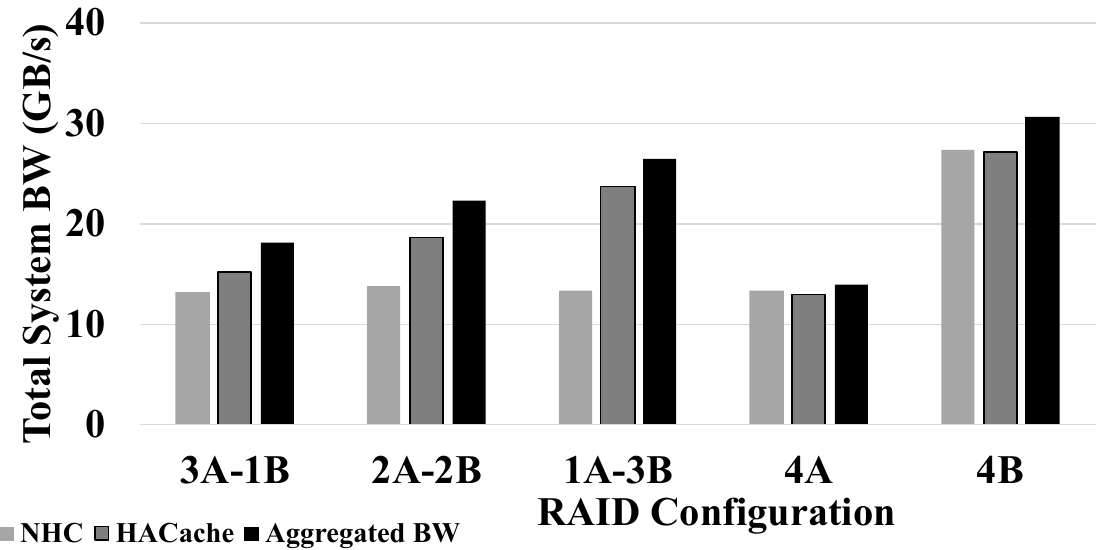}
  \caption{Performance comparison under sufficient cache capacity with 4KB random reads}
\label{fig:c4:eval:4kp} 
\end{figure}

\subsection{Evaluation under Deficient Cache Capacity}

To evaluate the effectiveness of the cache capacity regulation mechanism, we test HACache under scenarios with limited cache capacity.  
Since homogeneous RAIDs either have all backend SSDs with insufficient cache capacity or all with sufficient capacity, they are unaffected by cache regulation. Therefore, we only test heterogeneous RAID configurations “1A‑3B”, “2A‑2B”, and “3A‑1B”. In addition to NHC and aggregate bandwidth, we also evaluate HACache with cache capacity regulation disabled. Workloads include 4KB random reads and 128KB random reads. Cache size is set to 25\% of the I/O range, and fio is configured for uniform random reads. Without cache regulation, the hit rate is only 25\%.  

For comparison, the optimal diversion ratios required by the theoretical algorithm are as follows:  
- For 128KB random reads: {40\%, 40\%, 40\%, 0\%}, {50\%, 50\%, 0\%, 0\%}, and {55\%, 11\%, 11\%, 11\%}.  
- For 4KB random reads: {56\%, 56\%, 56\%, 0\%}, {66\%, 66\%, 0\%, 0\%}, and {73\%, 10\%, 10\%, 10\%}.  
In all cases, the actual hit rate is below the maximum required by the ideal diversion ratios.

\begin{figure}[htb]
    \centering
    \subfloat[128KB Random Read Performance Comparison]{
        \label{fig:c4:eval:128kd}
        \includegraphics[width=.45\columnwidth]{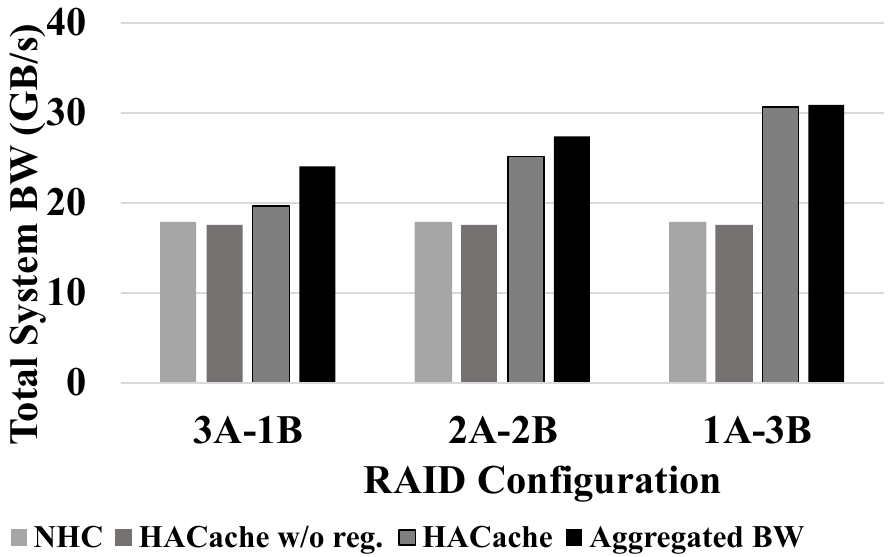}
    }
    \subfloat[4KB Random Read Performance Comparison]{
        \label{fig:c4:eval:4kd}
        \includegraphics[width=.45\columnwidth]{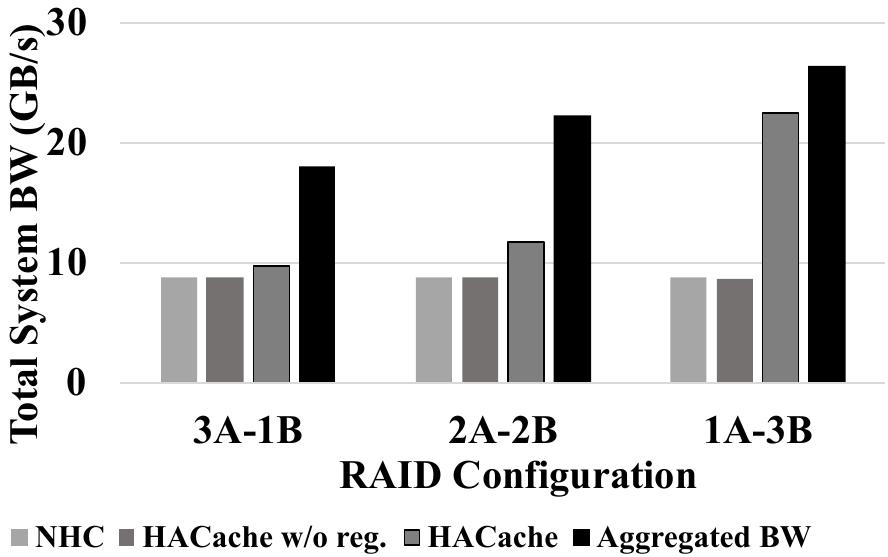} 
    }
    \caption{Read performance under limited cache capacity}
 \label{fig:c4:eval:deficient}
\end{figure}

Results are shown in Figures~\ref{fig:c4:eval:128kd} and \ref{fig:c4:eval:4kd}.  
For NHC, aggregate bandwidth utilization is only 74.7\%–58.0\% under 128KB random reads and 48.6\%–33.2\% under 4KB random reads. HACache without cache regulation performs similarly, since the cache SSD cannot effectively bridge the bandwidth gap between SSD A and SSD B. Although HACache without regulation utilizes SSD B’s bandwidth better than NHC, limited cache diversion ratios prevent full cache utilization, resulting in comparable performance.  

In contrast, HACache with cache regulation achieves 82\% to 99\% of aggregate bandwidth under 128KB random reads, demonstrating effective cache utilization. Under 4KB random reads, the “1A‑3B” configuration achieves 84.9\% of aggregate bandwidth, while “2A‑2B” and “3A‑1B” achieve only 52.7\%–54.0\%. This is because even with regulation, cache capacity allocated to slow SSDs is insufficient to provide high enough hit rates to support ideal diversion ratios. For example, in “3A‑1B”, regulation increases the hit rate for slow SSDs to about 30\%, far below the ideal 56\%. Nevertheless, even when ideal hit rates cannot be achieved, cache regulation still enables HACache to outperform the non‑regulated version.

\subsection{Sensitivity Analysis}
\subsubsection{Sensitivity of Convergence Time in Two-Phase Dynamic Valve Value Adjustment}

When the two-phase dynamic valve value adjustment strategy begins, the system accepts an initial set of valve values as the starting state. Clearly, the choice of initial state affects the optimization duration. In this section, we analyze the impact of initial states on convergence time. HACache’s two-phase dynamic valve value adjustment strategy collects bandwidth statistics per cycle, and at the end of each cycle updates valve values accordingly. To eliminate the influence of measurement duration, we use cycles as the unit of convergence time.  

For a storage architecture consisting of four RAID drives and one cache SSD, we performed a grid search over the initial valve vector space $\mathbf{P} \in [0,1]^4$ with a granularity of 0.01 to measure the number of cycles required for convergence. Due to the vast search space, actual testing is infeasible; instead, we designed a simulator based on the two-phase dynamic valve value adjustment strategy and system constraints (4.2)–(4.5), with step sizes set as in Section~\ref{sec:c4:imple}. We tested homogeneous and heterogeneous RAIDs under 4KB and 128KB random read workloads, assuming a 95\% cache hit rate.  

\begin{figure}[htb]
    \centering
    \subfloat[128KB Random Read Workload]{
        \label{fig:c4:eval:sen8}
        \includegraphics[width=.45\columnwidth]{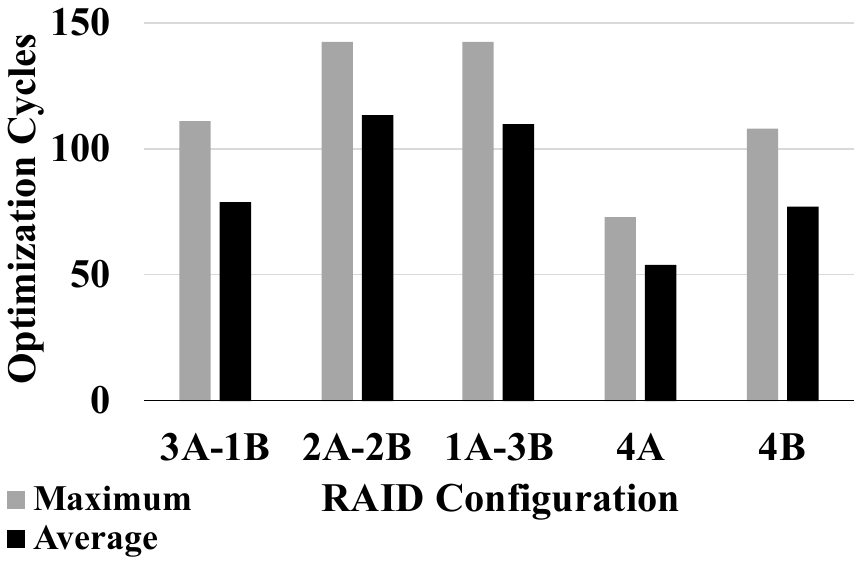} 
    }
    \subfloat[4KB Random Read Workload]{
        \label{fig:c4:eval:sen4}
        \includegraphics[width=.45\columnwidth]{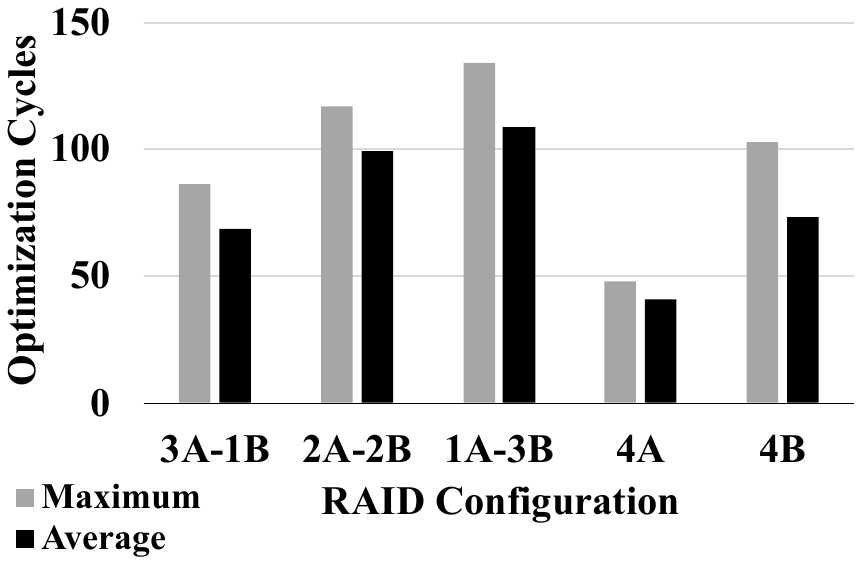}
    }
    \caption{Sensitivity of convergence time in two-phase dynamic valve value adjustment to initial configuration states}
 \label{fig:c4:eval:sen}
\end{figure}

We recorded average and maximum convergence cycles across all cases, as shown in Figure~\ref{fig:c4:eval:sen}. Overall, heterogeneous RAIDs require longer optimization times than homogeneous RAIDs, with average convergence cycles 1.56$\times$ higher. This is because in homogeneous RAIDs, all SSD valve values can be effectively optimized in each phase, whereas in heterogeneous RAIDs, only fast SSDs benefit while slow SSDs still undergo ineffective attempts, consuming time. Although these attempts have no effect under stable workloads, they can help adapt to workload variations in practice. Across all RAID configurations, maximum convergence cycles do not exceed 1.41$\times$ the average. Thus, while random initial values affect optimization speed, the system consistently converges to the optimal state within acceptable time.

\subsubsection{Sensitivity of Convergence Time in Capacity Regulation Strategy}

In this section, we analyze how cache hit conditions affect convergence time of the capacity regulation strategy. We define one iteration as consisting of cache warm-up, two-phase dynamic valve value adjustment, and cache capacity regulation (Figure~\ref{fig:c4:design:whexec}). To exclude variations in valve adjustment duration, we use iteration count as the unit of convergence time.  

We assume uniform random workloads and vary cache capacity from 1\% to 100\% of the I/O range to control hit rates. A simulator was developed to search this space, with step sizes set as in Section~\ref{sec:c4:imple}. We tested four-drive heterogeneous RAIDs under 4KB and 128KB random read workloads.  

\begin{figure}[htb]
    \centering
    \subfloat[128KB Random Read Workload]{
        \label{fig:c4:eval:sencap8}
        \includegraphics[width=.45\columnwidth]{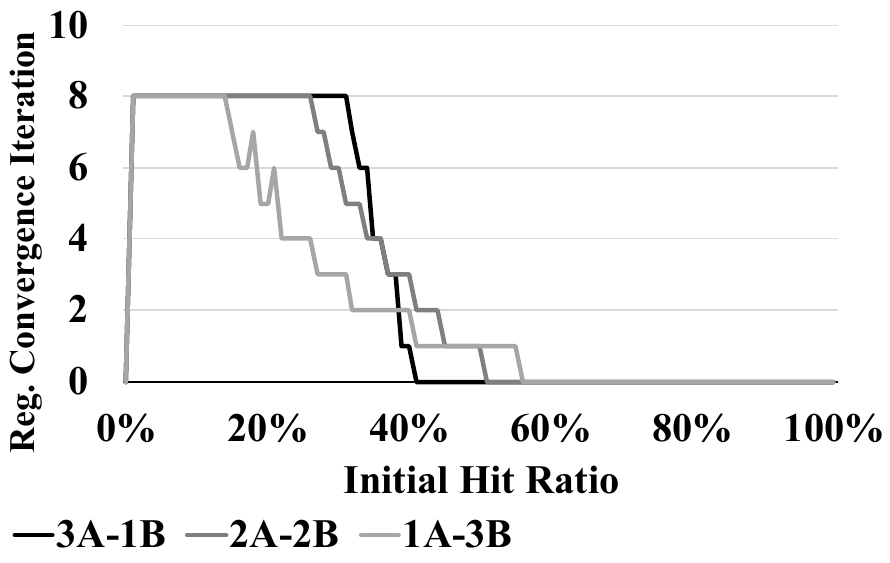} 
    }
    \subfloat[4KB Random Read Workload]{
        \label{fig:c4:eval:sencap4}
        \includegraphics[width=.45\columnwidth]{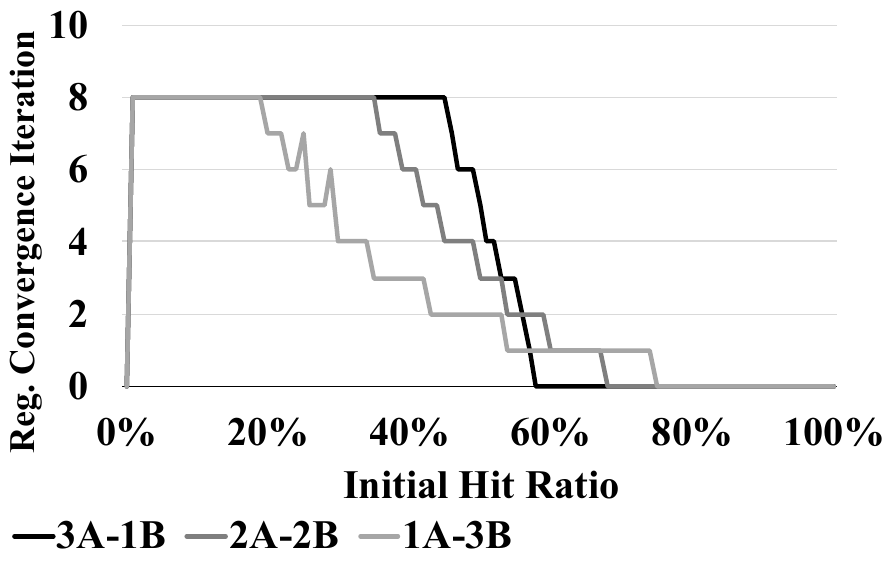}
    }
    \caption{Sensitivity of convergence time in capacity regulation strategy to cache hit conditions}
 \label{fig:c4:eval:sencap}
\end{figure}

We plotted convergence iteration curves for all cases, as shown in Figure~\ref{fig:c4:eval:sencap}. Overall, as cache capacity shrinks, more iterations are required. In our configuration, capacity regulation converges within 8 iterations. This is because, given the reclamation granularity, surplus SSDs can reclaim capacity at most 8 times; beyond that, oscillation occurs and regulation stops. Smaller reclamation granularity allows finer adjustments, but in practice may prevent the system from perceiving changes in valve values during two-phase adjustment, leading to stagnation and longer convergence times.

\section{Conclusion}
In this chapter, we propose a performance‑aware caching mechanism, \textit{HACache}, to improve bandwidth utilization in heterogeneous SSD arrays.  
The basic idea is to employ high‑performance SSDs as read caches, and dynamically regulate diversion ratios and capacity allocation to mitigate mismatches between request distribution and processing speed under striping. Specifically, we first define the optimization objective of cache diversion ratios and propose an optimal algorithm to compute the ideal diversion ratio for each backend SSD as a reference target. Then, HACache introduces a two‑phase dynamic diversion optimization strategy, enabling the system to automatically search for optimal diversion ratios at runtime. Finally, HACache adopts a cache capacity regulation strategy, dynamically adjusting cache allocation among SSDs based on diversion status and hit capability.  

Experimental results show that HACache significantly improves read performance of heterogeneous RAIDs. In typical mixed configurations, overall bandwidth is improved by an average of 35\%, validating the effectiveness of HACache in achieving a balance between performance and cost in practical deployments with constrained budgets.

\bibliographystyle{IEEEtran}
\bibliography{ref}




\end{document}